\documentclass[twocolumn,tighten]{aastex63}
\usepackage{amsmath}

\newcommand{\response}[1]{#1}

\newcommand\gpcyr{\, \rm Gpc^{-3}\,yr^{-1}}
\newcommand\msun{\, \rm M_\odot}
\newcommand\rsun{\, \rm R_\odot}

\begin{document}

\title{On the Mass Ratio Distribution of Black Hole Mergers in Triple Systems}

\correspondingauthor{Miguel A. S. Martinez}
\email{miguelmartinez2025@u.northwestern.edu}

\author[0000-0001-5285-4735]{Miguel A. S. Martinez}
\affil{ Department of Physics \& Astronomy, Northwestern University, Evanston, IL 60208, USA}
\affil{Center for Interdisciplinary Exploration \& Research in Astrophysics (CIERA), Northwestern University, Evanston, IL 60208, USA}

\author[0000-0003-4175-8881]{Carl L.~Rodriguez}
\affil{McWilliams Center for Cosmology and Department of Physics, Carnegie Mellon University, Pittsburgh, PA 15213, USA}

\author[0000-0002-7330-027X]{Giacomo Fragione}
\affil{ Department of Physics \& Astronomy, Northwestern University, Evanston, IL 60208, USA}
\affil{Center for Interdisciplinary Exploration \& Research in Astrophysics (CIERA), Northwestern University, Evanston, IL 60208, USA}

\begin{abstract}
Observations have shown that the majority of massive stars, progenitors of black holes (BHs), have on average more than one stellar companion. In triple systems, wide inner binaries can be driven to a merger by the third body due to long-term secular interactions, most notably by the eccentric Lidov-Kozai effect.  In this study, we explore the properties of BH mergers in triple systems and compare their population properties to those of binaries produced in isolation and assembled in dense star clusters.  Using the same stellar physics and identical assumptions for the initial populations of binaries and triples, we show that stellar triples yield a significantly flatter mass ratio distribution from $q=1$ down to $q\sim0.3$ than either binary stars or dense stellar clusters, similar to the population properties inferred from the most recent catalog of gravitational-wave events, though we do not claim that all the observed events can be accounted for with triples. While hierarchical mergers in clusters can also produce asymmetric mass ratios, the unique spins of such mergers can be used to distinguished them from those produced from stellar triples. All three channels occupy distinct regions in total mass-mass ratio space, which may allow them to be disentangled as more BH mergers are detected by LIGO, Virgo, and KAGRA.
\end{abstract}


\section{Introduction}
\label{sect:intro}

The LIGO/Virgo Collaboration has recently released the second Gravitational Wave Transient Catalog \citep[GWTC-2, ][]{lvc2020cat}. Together with results from the first two observing runs \citep[GWTC-1, ][]{lvc2019cat}, the full gravitational-wave (GW) catalog is comprised of more than $50$ events, and is transforming our understanding of black holes (BHs) and neutron stars (NSs) across cosmic space and time \citep{lvc2020catb,lvc2020catc}.  Despite this wealth of information, the origin of many of these binary mergers, particularly the binary BHs (BBHs) mergers, remains highly uncertain.

Various formation scenarios for BBH mergers have been proposed over the years, including isolated binary star evolution \citep[e.g.,][]{bel16b,demi2016,gm2018,Breivik2020}, dynamical formation in dense star clusters \citep[e.g.,][]{askar17,baner18,fragk2018,Rodriguez2018,sams18,ham2019,krem2019}, mergers in triple and quadruple systems \citep[e.g.,][]{antoper12,Silsbee2017,ll18,GrishinPerets2018,Rodriguez2018a,fragg2019,flp2019,fragk2019,liu2019,FragioneLoeb2020,MartinezFragione2020,VignaGomez2021,safar2020,2020ApJ...898...99H}, mergers of compact binaries in galactic nuclei \citep[e.g.,][]{GondanKocsis2018,rasskoc2019,GondanKocsis2021}, mergers of binaries induced by a central MBH \citep[e.g.,][]{Hoang2018,steph2019,Wang2020}, mergers in active galactic nucleus disks \citep[e.g.,][]{bart17,sto17,mck2020}, and mergers of primordial BHs \citep[e.g.,][]{sasaki2016}. While several formation scenarios can account for some or even all of the BBH merger rate \citep{2021RNAAS...5...19R}, the contribution of different channels must be disentangled using a combination of the mass, spin, redshift, and eccentricity distributions as the number of detected events increases \citep[e.g.,][]{olea09,Fishbachetal2017,gondan2018,perna2019,WongBreivik2020,zevin+2021}.

The majority of the BBHs in GWTC-2 are consistent with having equal mass components \citep{lvc2020catb}. This is also consistent with theoretical models of BBH formation through either isolated binary evolution or dynamical formation, which prefer BBH mergers with mass ratios near unity \citep[e.g.,][]{bel16b,Rodriguez2016}. Nevertheless, GW190412, a BBH with mass ratio of nearly four-to-one, provides evidence for the existence of mergers with small mass ratios. It has been proposed that these systems can be formed through isolated binary evolution \citep[e.g.,][]{OlejakFishbach2020} or as a result of repeated mergers in star clusters \citep[e.g.,][]{2020ApJ...896L..10R}.

Another channel that can potentially produce BBH mergers with smaller mass ratios is the Lidov-Kozai (LK) mechanism and the associated eccentric LK mechanism in triple systems \citep{vonZeipel1910,Kozai1962,Lidov1962,Naoz2011}. In this scenario, a tertiary companion on a sufficiently inclined outer orbit could drive the inner binary to extreme eccentricities, leading to efficient GW emission and orbital decay. In this study, we examine in detail the mass ratio distribution of BBH mergers induced by tertiary companions in triple systems and compare them to the mass ratio distributions from detailed binary population synthesis and star cluster dynamics. We show that the triple channel predicts a flatter mass ratio distribution than either isolation binary or dynamical formation scenarios, and occupies a distinct region of the of total mass, mass ratio, and spin parameter space for merging BBHs, which can be used to disentangle the triple contribution to the overall observed GW sources.

The paper is organized as follows. In Section~\ref{sect:methods}, we present our method to build populations of merging binary BHs. In Section~\ref{sect:results}, we compare the distributions of total masses and mass ratios for different channels. Finally, in Section~\ref{sect:concl}, we discuss the implications of our results and draw our conclusions.

\section{Models of Merging Binaries}
\label{sect:methods}

We compare three populations of merging BBHs: those created from mergers from ``isolated'' stellar binaries, those from stellar triples where the merger is induced by the third companion due to secular interactions, and those formed through three- and four-body encounters in dense star clusters.  As with any population synthesis study, the choice of initial conditions and physical choices could greatly effect the resulting distributions of BBH mergers, including the shapes of different distributions and the expected rates of events \citep[e.g., the range of models presented by][]{bel2020}.  This issue is even more apparent given the wide range of possible outcomes for a single binary system depending on the assumptions made \citep{2021arXiv210810885B}.  As such, any claims regarding the difference between formation channels must be made using \emph{the same} initial conditions and physical choices.  As much as possible, the three sets of results presented here strive to use identical prescriptions for the evolution of massive stars.  To that end, both the binary and triple evolution are generated using the \texttt{COSMIC} software package for population synthesis \citep{Breivik2020}.

Below, we list the key physics of binary stellar evolution and the relevant assumptions we choose for each.
\begin{itemize}
    \item Mass transfer: In a binary system, it is possible for a star's radius to increase such that it fills its Roche Lobe. When this happens, it transfers mass onto the other star via Roche Lobe Overflow (RLOF). In the case of stable mass transfer, we calculate the rate of mass loss from the donor star following \cite{Hurley2002} and we assume that mass transfer onto the accretor is limited by the Eddington rate for compact objects. For non-compact objects, the accretion rate during Roche Lobe overflow is $10$ times the thermal rate of the accretor for Main Sequence, Hertzsprung Gap, and Core Helium Burning stars and unlimited for Giant Branch, Early Asymptotic Giant Branch, and Asymptotic Giant Branch stars. We assume that any material from the system is lost with the specific angular momentum of the secondary.
    
    \item Common Envelope: Under certain conditions, mass transfer can become unstable. When this happens, the two stars can inhabit a common envelope, subjecting both stars to viscous drag forces and shrinking the binary separation to very small values. To determine the critical mass ratios for the onset of unstable mass transfer, we follow \cite{Belczynski2008} except when the donor is a white dwarf, for which we follow \cite{Hurley2002}. To determine the outcome of the common envelope phase, we use the common ``$\alpha\lambda$'' prescription, where $\alpha$ describes the efficiency of transferring orbital energy to the envelope and $\lambda$ describes the binding energy of the envelope \citep{vandenheuvel76,Webbink1984,livio+soker88,dekool90,iben+livio93,dewi+tauris2000,dewi+tauris2001}. For this study, we set $\alpha=1$ and we use the variable $\lambda$ prescription of \cite{Claeys2014}, without extra ionization energy. Finally, when at least one member of the binary is a star without a clear core--envelope boundary, we use the ``optimistic'' prescription, which allows the common envelope phase to proceed normally for these systems, as opposed to the ``pessimistic'' prescription, which assumes that such episodes always lead to a merger within the common envelope.
    
    \item Remnants: We model the remnant masses following SN with the ``rapid'' prescription of \cite{fryer2012} in order to be consistent with the assumptions of the CMC models. The default prescription used by \texttt{COSMIC} is the ``delayed'' prescription of \cite{fryer2012}, which allows for the formation of objects between in the range $[3-5]\,\msun$ (low-mass gap).
\end{itemize}

The input assumptions (both initial conditions and physical prescriptions) chosen for this paper for the field channels are identical to those of the fourth model of Table 1 from \cite{Zevin2020}. For a full description of the rest of the default \texttt{COSMIC} prescriptions, please refer to \cite{Breivik2020} and the \texttt{COSMIC} documentation.

We describe our choices for initial conditions for the different formation channels in the sections below.

\subsection{Binary Black Holes from Stellar Binaries and Triples}

The typical massive star (those that are the progenitors of stellar-mass BHs) has $\gtrsim 2$ companions \citep{Sana2012, Moe2017}, suggesting that any population synthesis studies of binary evolution should consider many of those binaries as subsets of a larger triple (and other multiple star system) population.  As such, our approach here uses a single population of triples as the basis for \emph{both} the binary and triple population synthesis. 

For each triple, we begin by drawing a primary mass from a \cite{Kroupa2001} initial mass function (IMF) following a $p(m) \propto m^{-2.3}$ power law from 18$M_{\odot}$ to 150$M_{\odot}$, roughly reflecting the interval of BH-progenitor masses.  Our choice of inner and outer binary periods are informed by Figure 2 of \cite{Moe2019}: the inner binary periods and eccentricities are sampled from \cite{Sana2012}, consistent with several previous population synthesis studies of massive binaries \cite[e.g.][]{Zevin2020,OlejakFishbach2020}, while the outer periods are sampled from a log-normal distribution centered at $10^{3.8}$ days with a width of $\sigma_{\log P}=2.4$ and an upper limit at $10^{6.5}$ days. For each triple we redraw outer orbital periods until $P_{\rm inner} < P_{\rm outer}$, and draw the eccentricities of the outer orbit from a thermal distribution \citep[$p(e)\propto e$,][]{Ambartsu1937,Heggie1975}.  The mass ratio between both the primary/secondary and primary/tertiary stellar pairs are sampled from the joint mass-ratio/period distributions presented in \cite{Moe2017}, using the marginalized $q$ distributions for each primary mass, inner, and outer periods.  Once we have masses, periods, and eccentricities for each triple, the geometric alignment of the two orbits is sampled isotropically in the relevant angles: uniform in the cosine of the mutual inclination, and uniform in the angle of the pericenter and longitude of the ascending nodes of both the inner and outer binaries.

Once we have our initial population of stellar triples, we must integrate them forward until we are left with a population of BH triples and binaries.  We evolve each triple forward for 100 Myr until each star has collapsed to form a BH or a neutron star (which are then discarded). Since \texttt{COSMIC} only evolves single or binary stars, we model these triples using the same approach as \cite{Rodriguez2018a}: the inner binary and tertiary companion are initially evolved as isolated systems until all three stars have collapsed into compact objects. Thus, we assume implicitly that all the triples we study are non-interacting, such that there is no mass transfer from the tertiary to the inner binary. However, we still consider interactions \textit{within} the inner binary, i.e. mass transfer. While it is possible for the tertiary to fill its Roche Lobe and transfer mass onto the inner binary \citep[e.g.][]{deVries+2014,portegieszwart+leigh2019,distefano2020,Leigh+2020,comerford2020,Toonen2020,glanz2021,Hamers2021}, \response{assuming there is no mass transfer from the tertiary to the inner binary} is valid for almost all the triples we consider in our study. Indeed, using initial conditions similar to those used here, \cite{Toonen2020} and \cite{Hamers2021} have recently shown that only a very small fraction ($\sim0.5-1\%$ for the former, $<0.1\%$ for the latter) of \textit{all} triples could undergo mass transfer from the tertiary onto the inner binary.

At each timestep during the evolution of both the inner and the outer orbit, the mass loss from the stars and supernovae--as well as their associated natal kicks--are allowed to modify the orbital periods, eccentricities, and geometric orientation of the triples.  The orbital periods expand adiabatically due to stellar mass loss, while the orbital shapes and orientations change due to the velocity kicks from supernova.  For our population of binaries we only consider those systems that produce a bound BBH that will merge within the age of the universe (regardless of whether the third companion is bound) due to GW emission alone.  We do not impose this requirement for our population of triples, since the third object may drive the inner binary to merge. However, we do require that all three objects remain bound, and are sufficiently hierarchical for the LK mechanism to be relevant: $P_{\rm outer}$ must be sufficiently large that the triples are dynamically stable, but not so large that relativistic precession of the inner binary can quench the secular influence of the third object \citep[following][]{Antonini2018}.  We throw out any systems that become dynamically unstable \citep[using the criterion from][]{Mardling2001} at any point during their integration.  

We integrate our BH triples dynamically using the vector form of the three-body secular equations of motion expanded to octupole order \cite[see][]{Liu2015}.  We also include the equations of motion describing the pericenter precession of the inner binary, the spin-orbit and spin-spin coupling of the inner BH spins (using the initial spin tilts of the post-supernova binaries), and GW emission for the inner binary.  See \cite{Rodriguez2018a} for details.  

We integrate $10^6$ triple systems for 11 different stellar metallicities (Z of 0.0001, 0.00025, 0.0005, 0.00075, 0.001, 0.0025, 0.005, 0.0075, 0.01, 0.02, and 0.03).  Each system is then assigned a cosmic birth time following the star formation rate \citep{Madau2014} convolved with the chemical enrichment prescriptions of \cite{Belczynski2016}.  The latter assumes that the metallicity of stars forming at redshift $z$ is given distributed according to a log-normal distribution, with a given mean and standard deviation of 0.5 dex \cite[see][]{Belczynski2016,Dvorkin2015}.  Each of the 11 stellar metallicities we consider is assigned all the star formation that occurs within that metallicity bin, and the birth times are sampled accordingly.Following BH formation, we calculate the merger time due to GWs \citep[][]{Peters1964} for all inner binaries that remain bound, including systems where the tertiary became unbound. For all triples that remain stable and bound, we integrate them as described above in the case the inner binary's GW merger timescale is greater than 13.8 Gyr. This provides us with a population of binary and triple BBH mergers sampled by using the same stellar physics and cosmic star formation history.  Both populations were integrated for 13.8 Gyr, making it possible that a given binary or triple may merge after the present day (given a sufficiently late birth time).  We exclude these systems from our analysis.  

\subsection{Binary Black Holes from Globular Clusters}
\label{sec:gcs}

Our population of globular cluster (GC) BBHs are taken from a grid of $N$-body models of star clusters \citep{Kremer2020}.  These models, developed using the Cluster Monte Carlo code \cite[CMC,][]{Joshi2000,2013ApJS..204...15P} contain all the necessary physics to self-consistently describe the evolution of dense spherical star clusters, such as globular and super star clusters, over many relaxation times.  \texttt{CMC} includes dynamical processes such as two-body relaxation \citep{Henon1971}, three- and four-body dynamical encounters between stars, BHs, and binaries, and (critically for this study), single and binary stellar evolution using nearly identical prescriptions from \cite{Hurley2000,Hurley2002,Breivik2020} that are used in \texttt{COSMIC}.  While there have been minor changes to the prescriptions for binary evolution, particularly those related to mass transfer and the common envelope phase, between the version of \texttt{COSMIC} used here and the assumptions in the grid of \texttt{CMC} models, the underlying single star evolution is nearly identical, and it has been shown \citep{Rodriguez2016} that binary physics does not significantly influence the properties of dynamically-assembled BBHs from clusters.  The only significant change to the evolution of massive stars is a modification to neutrino emission for collapsing BHs at the lower end of the pulsational-pair instability mass gap, which increases the maximum BH mass for single stars from $40.5M_{\odot}$ to $45M_{\odot}$.

To place the dynamically-formed BBHs in a cosmological context, we follow a similar procedure to \cite{Rodriguez2018c}.  We bin the 148 cluster models in equally spaced bins in cluster mass and logarithmically-spaced bins in metallicity, and assign to each model a weight corresponding to the number of GCs that are thought to form in that 2D bin across cosmic time.  The masses follow a $p(M_{\rm GC}) \propto 1/M_{\rm GC}^2$ cluster IMF, while the metallicities are taken from the GC formation model of \cite{ElBadry2019}.  Similarly, the formation time of each BBH-host cluster is drawn from the distribution of cluster formation times for a given metallicity, which is then added to the merger time for that BBH within each cluster model.  See \cite{Rodriguez2018c} for details. Systems that merge later than the present day are discarded from our analysis.

\subsection{Detectability}

In addition to the astrophysical populations of merging BBHs, we are also interested in the population of BBHs detectable by advanced GW observatories.  Here we follow the technique presented in \cite{Rodriguez2019}.  We assign each BBH a detectability weight defined as   

\begin{equation}
    w_{\rm det} \equiv f_d (m_1,m_2,z) \frac{dVc}{dz} \frac{dt_s}{dt_o} ~,
    \label{eqn:gwweight}
\end{equation}

\noindent where $f_d$ is the fraction of sources with masses $m_1$ and $m_2$ merging at redshift $z$ that are detectable (defined as a signal-to-noise ratio of 8 or higher) by a three-detector network operating at the design sensitivity of Advanced LIGO \citep{Abbott2018}.  $\frac{dVc}{dz}$ is the amount of co-moving volume in a slice of the universe at redshift $z$, and $\frac{dt_s}{dt_o} \equiv \frac{1}{1+z}$ is the difference in co-moving time between the merger redshift and the observer at $z=0$.  We use the IMRPhenomD GW approximant for computing GW detectability \citep{2016PhRvD..93d4006H,2016PhRvD..93d4007K}. Each of the detectable quantities presented here are weighted using Eq. \eqref{eqn:gwweight}.

\section{Results}
\label{sect:results}

\begin{figure*}
\includegraphics[width=0.96\textwidth]{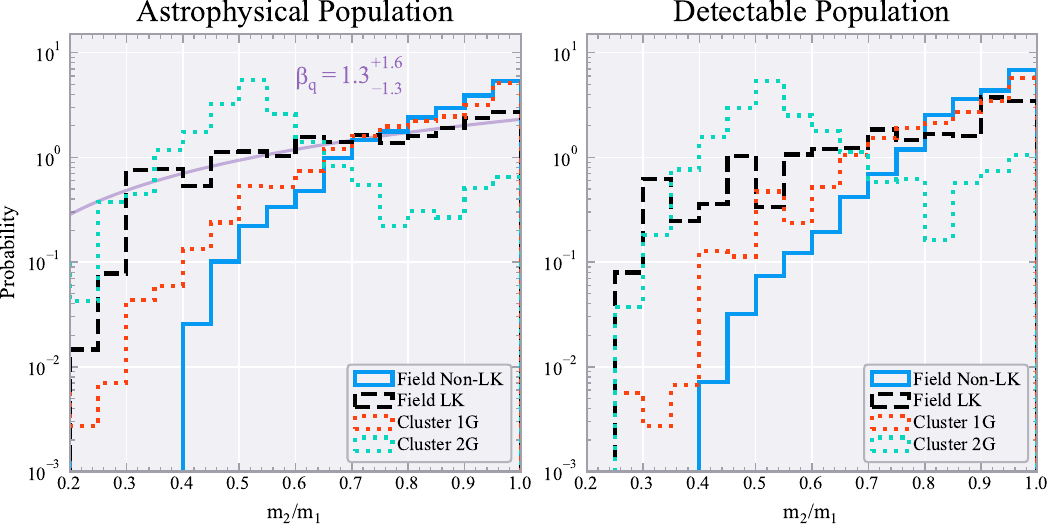}
\caption{Distribution of mass ratios for different BBH formation channels. The merging sample is weighted by star (cluster) formation history for the field (cluster) mergers (left) and detectability (right), to produce realistic astrophysical and detectable populations in the local universe. In addition to the field non-LK (solid blue) and field LK (dashed black) distributions, we also include the mass ratio distributions from GCs using data presented in \cite{Kremer2020}, split into 1G (dotted orange) and 2G (dotted green) channels. The median for the mass ratio distribution from the Power-law model from GWTC-2 (solid purple line) is presented in the left panel, normalized such that the power law integrated from 0.1 to 1 sums to 1.}
\label{fig:massratio}
\end{figure*}

\begin{figure*}
\centering
\includegraphics[]{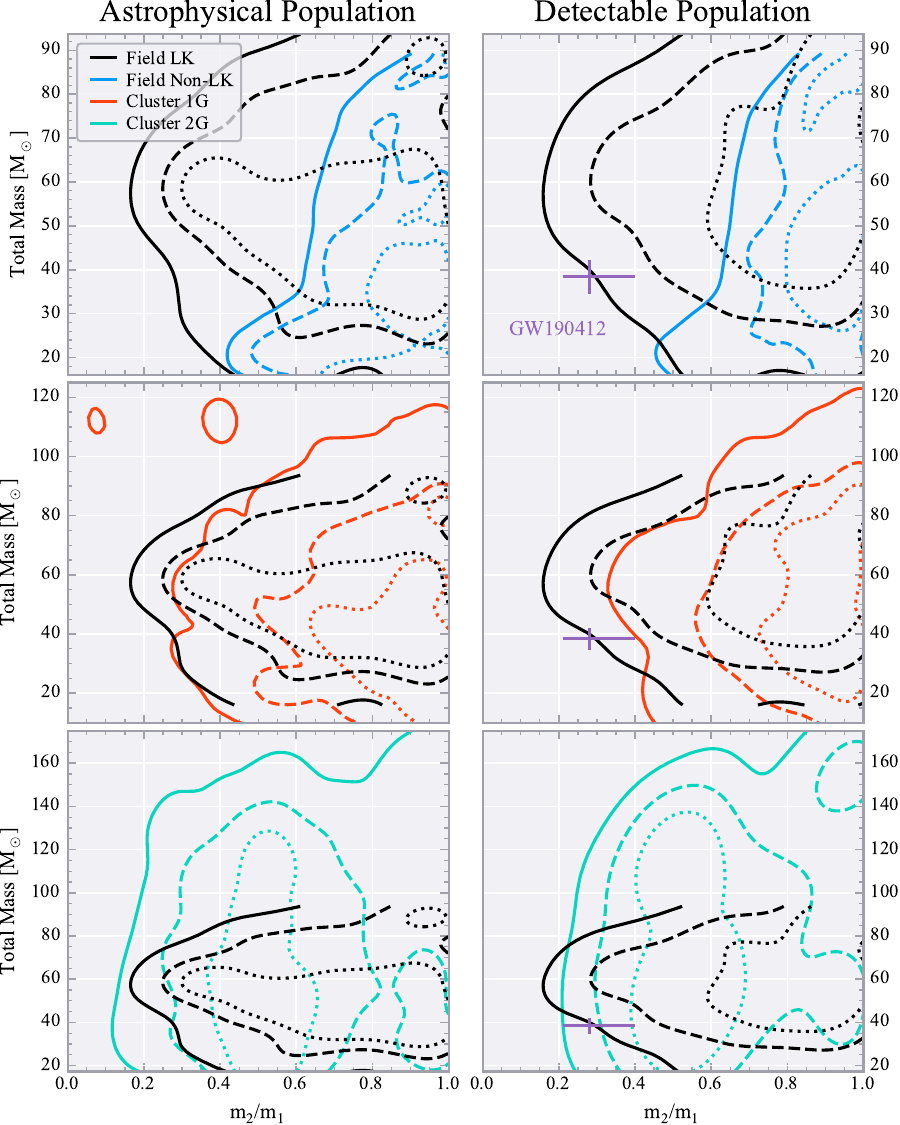}
\caption{Contour plots for mass ratio--total mass distributions for different channels. The contours show the regions containing 99.5 (solid), 90 (dashed), and 60 (dotted) percent of all data points. The left column shows astrophysically weighted samples while the right column shows the same samples weighted by detectability. In each row, the field LK channel (black) is compared to the field non-LK (top, blue), cluster 1G (middle, orange), and cluster 2G (bottom, green) channels. Also included in the right column is the detection of GW190412 during O3a \citep{theligoscientificcollaboration2020}. Note the differing range in the y-axis of each row. In Field populations, the total mass of the merging binary is limited by pulsational pair instability. However, in the Cluster populations, the maximum birth mass of a BH from stellar collapse can be over this limit because of repeated stellar mergers prior to stellar collapse.}
\label{fig:mtot_q}
\end{figure*}

\begin{figure}
\includegraphics[]{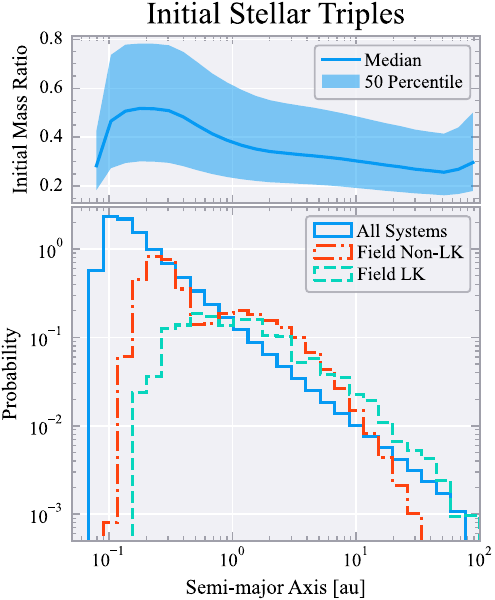}
\caption{Top: median and 50 percentile initial inner binary mass ratio as a function of inner semi-major axis. Bottom: normalized distribution of inner semi-major axis prior to stellar evolution for all initial triples in the sample, \response{including those that didn't result in a merger} (solid blue), merging binaries (dash-dotted red), and merging triples (dashed green).}
\label{fig:a1i}
\end{figure}

\begin{figure}
\includegraphics[]{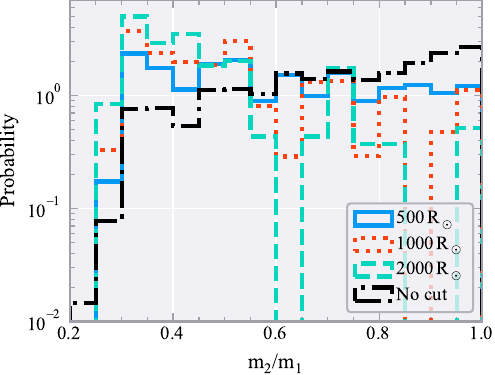}
\caption{Distribution of mass ratios for different cuts to the triple sample. The dash-dotted black line matches that of the left panel of Figure \ref{fig:massratio}. The solid blue, dotted orange, and dashed green lines represent the sample where all stars which may have experienced LK-driven Roche Lobe overflow are discarded for different assumed values of maximum stellar radius.}
\label{fig:rlof}
\end{figure}

For the purposes of comparison between the different channels, we delineate between Field Non-LK, Field LK, Cluster 1G, and Cluster 2G mergers. The Field Non-LK and Field LK mergers are produced using the procedure described in the previous section. We designate the Field LK (LK) mergers as those taking place in useful triples, which required the presence of the outer companion for the inner binary to merge via the eccentric Kozai-Lidov effect. The Field Non-LK (non-LK) are those inner binaries that could merge within the age of the Universe solely due to GWs, regardless of whether or not the outer companion was still present post-BH formation or was previously unbound from the inner binary due to stellar evolution or natal kicks. Thus, we consider what we call the Field Non-LK channel here as equivalent to the classic isolated binary evolution channel. As we discuss below, counting the non-useful triples as part of the Field LK channel instead of the Field Non-LK channel, which we do here, does not change much. From our population synthesis of $11\times10^6$ triples we find 858 ($7.8\times10^{-3}\%$) LK mergers and 135104 ($1.2\%$) non-LK mergers. Cluster 1G mergers are dynamically-assembled mergers which involved only first-generation BHs (those created from stellar collapse), whereas Cluster 2G mergers involve at least one second-generation BH. We do not distinguish between LK and non-LK mergers in clusters, as the triple companion was shown to have minimal effect on both the mass and mass ratio distribution \citep{MartinezFragione2020}. 

In Figure \ref{fig:massratio}, we compare the mass ratio ($q=m_2/m_1$ for $m_1>m_2$) distributions of BH mergers events from the four populations. On the left side, we show the field (cluster) population weighted by star (cluster) formation history. On the right, we further weight the populations by detectability. On the left side, we also show the power law model (with an arbitrary normalization) for $q$ from the \textsc{Power Law + Peak} model ($\beta_q=1.3^{+1.6}_{-1.3}$) fits from GWTC-2 \citep{lvc2020catb} with the purple line and shaded purple band. Using the \textsc{Broken Power Law} model ($\beta_q=1.4^{+1.7}_{-1.4}$) instead of the \textsc{Power Law + Peak} model gives nearly identical results.

We find that the mass ratio distribution for the field non-LK mergers here agree well with previous results \response{from isolated binaries} \citep[e.g.,][]{Zevin2020}. The cluster 1G and 2G distributions are both shaped by binary-mediated dynamical interactions in the cores of dense stellar systems, which will typically exchange lower-mass components of binaries in favor of producing equal mass BBHs \cite[e.g.,][]{1993ApJ...415..631S}. Thus, there is a preference for equal mass mergers in the 1G case and for mergers with $q\sim0.5$ in the 2G case, as the heaviest BH in the cluster, typically the remnant of a 1G merger, will be roughly twice the mass of the next heaviest BH. Remarkably, we find that the mass ratio distribution for LK mergers is largely flat between roughly 0.3 and 1.0.  This is particularly true compared to non-LK mergers, which means that--given the same initial conditions--the LK mechanism preferentially enhances the merger rate of lower mass ratio systems. \cite{2021arXiv210301963S} demonstrated that octupole-order effects in hierarchical triple evolution cause the merger fraction to increase as the mass ratio of the inner binary is decreased. We also find that the inner binaries of the LK mergers go through different evolutionary pathways than the non-LK mergers. In our sample, only a single LK system experienced a common envelope phase, whereas the majority of the non-LK systems ($77.6\%$) required a common envelope phase in order to merge within a Hubble time \citep[][]{bel16b,demi2016,gm2018,Breivik2020}. Just as previous studies have shown, when the inner binary of a triple undergoes a common envelope phase, the LK effect is quenched by GR precession \citep[e.g., ][]{Antonini2012,Rodriguez2018a}. On the other hand, the inner binary of LK systems typically evolve only through episodes of stable mass transfer ($94\%$ for LK mergers, compared to $22.4\%$ for non-LK mergers), which tend to produce more extreme mass ratios and larger separations, thus requiring the LK cycles imposed by the tertiary in order to merge within a Hubble time \citep[][]{Toonen2020}. As a result, the mass ratio distribution produced through this channel produces low mass ratio mergers compared to the isolated binary channel when using self-consistent initial conditions. Furthermore, we find that this mass ratio distribution is the one that most closely tracks the inferred distribution function from GWTC-2 \citep{lvc2020catb}. Going from the astrophysical population to the detectable population, many of the same general trends are present, though all the distributions are slightly biased towards equal mass ratios.

In Figure \ref{fig:mtot_q}, we present the same mass ratio distributions as above against the total binary masses of each merging system as contour plots, where we show the regions containing 60\% (dotted), 90\% (dashed), and 99.5\% (solid) of the sample. The left column shows the astrophysical population and the right shows the detectable population. In each row, we compare the field LK populations (black contours) to the field non-LK (blue contours), cluster 1G (orange contours), and cluster 2G (green contours) populations. We also include in the right column in each row GW190412 \citep{theligoscientificcollaboration2020}. 

Many of the general features from Figure \ref{fig:massratio} are also present here, such as the large number of LK mergers with low mass ratios. In the top row, we show that when comparing mergers with $q\lesssim0.5$, LK mergers tend to have a higher total binary mass compared to the non-LK mergers, with the center of the total mass ratio distribution at $\sim50-60\msun$ for LK mergers, whereas non-LK mergers appear to always have total mass less than $\sim40\msun$. In the middle row, this observation still roughly holds when comparing the field LK and cluster 1G mergers, where it is clear that the mergers produced by the LK channel tend to have a lower mass ratio. Finally, the bottom row demonstrates that the peak of the cluster 2G distribution nearly occupies the same portion of $M_{\rm tot}-q$ space as the field LK mergers. However, note that if BHs are born with negligible spin \citep{Fuller2019a}, as is assumed in this study and in the GC models \citep{Kremer2020}, then these two populations will be easily distinguished by the spins of the components, since the primary 2G BH should have a spin $\chi\sim0.7$. If BHs are born with significant ($\chi \gtrsim 0.2$) spins from stellar collapse, then the relative contribution of hierarchical mergers from GCs will be greatly diminished \citep[e.g.,][]{Rodriguez2019,FragioneLoeb2020b}. Finally, we observe that GW190412 is consistent (with $99\%$ confidence) with either an isolated LK or 2G origin based solely on its position in $m_{\rm tot}-q$ space. The detection announcement of GW190412 also reported a confident detection of the spin of the primary BH $\chi_1=0.44_{-0.22}^{+0.16}$ while the spin of the secondary BH is largely unconstrained \citep{theligoscientificcollaboration2020}. If BHs are born with high spin, then by the logic above, our models would suggest an isolated LK origin is a more likely explanation between the two.

As stated previously, we found that the majority of merging non-LK systems undergo at least one common envelope phase, whereas only $0.1\%$ merging LK systems do. In the top panel of Figure \ref{fig:a1i}, we show the median and 50 percentile of the initial mass ratio distribution of our sample as a function of semi-major axis. We show in the bottom panel the distribution of ZAMS inner semi-major axis for our total initial sample (solid blue), as well as the ZAMS inner semi-major axis of the systems that merge without LK (dash-dotted red) and with LK (dashed green). In the bottom panel, it can be seen that the two merging populations sample different areas of the initial inner semi-major axis distribution, which also have different initial mass ratios. Whereas the \response{non-LK mergers} tend to have smaller initial semi-major axes, with a very apparent enhancement at $\sim0.1\,\rm{au}$, the LK mergers have a much larger range of initial semi-major axes. Accordingly, the merging LK systems tend to have smaller initial mass ratios than the non-LK systems. These different mass ratio distributions are then sculpted by further binary stellar evolution and these subtle differences are further amplified, most notably through the presence or absence of common envelope phases.

In this study we considered only the useful triples. In principle, it is possible that the ``non-useful'' triples, or those whose inner binary would have merged regardless of the tertiary companion, could still enhance the rate of triple mergers within the local Universe. We find that $15.3\%$ of the non-LK mergers take place within a stable triple system. However, only $0.03\%$ of these mergers take place in a triple system in which the LK mechanism would not be quenched due to relativistic pericenter precession \citep[see also Figure 4 of][]{Rodriguez2018a}. As a result, we expect that these non-useful triples would account for at most an additional few percent of LK mergers where the LK mechanism allows for a substantial decrease in the delay time. For this reason, we count the non-useful triple mergers as part of the Field Non-LK mergers since they are indistinguishable from a non-LK merger. Furthermore, in this study we only considered stable triples where all three components are BHs. We found that, for triples (both useful and non-useful) where the inner binary was composed of two BHs, about $2/3$ of the triples had a BH as the outer companion. While a detailed treatment is outside of the scope of this work due to the possibility of mass transfer from the tertiary \citep[e.g.][]{deVries+2014,portegieszwart+leigh2019,distefano2020,Leigh+2020,comerford2020,Toonen2020,glanz2021,Hamers2021}, assuming a comparable merger fraction for the non-BH companions leads to a roughly 50\% increase in the number of triple mergers.

While our approximate technique for modeling hierarchical triples can track the evolution of the triple configuration due to mass loss, supernova, and the evolution of the inner binary, it does not allow for the possibility of LK oscillations prior to BH formation. As a result, it is possible that many of the systems presented here may actually have merged prior to their formation as BH triples. However, we can estimate the effect this missing physics may have on our results. In order to explore the sensitivity of our results to LK oscillations during the main sequence and giant phases, we calculated the minimum pericenter distance of each of the inner binaries of the triples attainable due to LK \citep[by numerically solving Eq.~2 of][]{Liu+2019} from their zero-age main sequence initial conditions. We then compared these separations to the distance required for each star to fill its Roche lobe \citep[Eq.~2 of][]{Eggleton83}, assuming all the stars acquired an average maximum radius of $500$, $1000$, and $2000\rsun$ \citep[e.g.,][]{Agrawal+2020}, and discarded any systems that would have passed within their Roche lobe separation. We find that $162\,(19.3\%)$, $84\,(9.8\%)$, and $33\,(3.8\%)$ of the systems survive these cuts for each assumed radius respectively. Note that for stars with ZAMS mass $100\msun$, the \texttt{SSE} fitting formula for stellar radii employed by \texttt{COSMIC} produce maximum radii nearly as large as $10^4\rsun$. If this value is assumed, then no systems would survive this cut. However, the results of \cite{Agrawal+2020} suggest that the smallest assumed value, $500\rsun$, is a more realistic average maximum radius across the entire mass spectrum. The resulting change to the mass distribution is illustrated in Figure \ref{fig:rlof}, where the additional constraint results in a distribution even more biased to lower mass ratios, and this effect increases when assuming larger maximum radii. This is to be expected, since excluding systems which may have undergone binary interactions or mergers will also preferentially select the ultra-wide triples illustrated in the right side of Figure \ref{fig:a1i}. The initial inner binary semimajor axis of the widest triples in our merging sample can have inner binaries at separations between $10$ and $100$ AU, decreasing the likelihood of any mass transfer occurring, even in the presence of LK oscillations. The top panel of Figure \ref{fig:a1i} also shows that these same inner binaries tend to have smaller initial mass ratios, so that our cuts to the sample are preferentially selecting low mass ratio systems.

\section{Conclusions}
\label{sect:concl}

In this paper, we started from a sample of triple massive stellar systems and evolved them to their end state as BH triples, in order to obtain a self-consistent comparison between the binary and triple channels. We also compared this population with a population of mergers from previous studies studying BBHs in dense stellar clusters which used the same assumptions for stellar evolution \citep{Kremer2020}. 

We report that the mass ratio distribution from the isolated triple systems in our sample produce a BBH mass ratio distribution that is nearly flat from unity to $\sim0.3$, producing significantly more low-mass-ratio mergers than either isolated binary evolution of dynamical formation of BBHs, consistent with \cite{2021arXiv210301963S}. Furthermore, we show that the triple channel seems to agree with the power law implied by the most recent catalog of GW observations better than other channels. As a result, the triple channel may be able to more efficiently produce events with low mass ratios such as GW190412.

However, our results do not suggest that mergers induced by LK are able to reproduce the whole catalog of observed events. We simply assert that if isolated BH triple systems are produced in nature, then we may be able to gauge the contribution of the LK mechanism to the total population of merging BH systems by low mass ratio and high total mass. This is in addition to other means of distinguishing LK mergers, such as through eccentricity in the LIGO band or effective spin distribution \citep[e.g.,][]{Antonini2012,Antonini2017,Liu2017,ll18,Rodriguez2018c,fk2020}.

Apart from the mass ratio distributions, it is important to quantify the relative contribution of the LK channel to the total BBH merger rate.  While a complete estimate of the BBH merger rate is beyond the scope of this study, we can approximate the rate by multiplying the ratio of LK BBH mergers to binary BBH mergers using our initial conditions ($0.003$) by the volumetric merger rate at $z=0$ using the same physics and (binary) initial conditions from the literature \citep[$\sim 1.6\times10^3 \gpcyr$,][]{Zevin2020}.  This suggests a local LK merger rate of $\sim5 \gpcyr$, consistent with previous estimates of $0.02-25 \gpcyr$ \citep[e.g.][]{Silsbee2017,2017ApJ...841...77A,Rodriguez2018a,flp2019}.  The initial conditions and common envelope physics used in \cite{Zevin2020} clearly produce extremely unrealistic volumetric rates for BBH mergers (up to two orders of magnitude larger than the observed rate of $23.9^{+14.3}_{-8.6} \gpcyr$ \citep{lvc2020catb}), meaning our ratio of LK to non-LK mergers may similarly be a significant underestimate. For comparison, the BBH merger rate predicted for the clusters presented in this work is roughly $17 \gpcyr$ \citep[][]{2021RNAAS...5...19R}\footnote{This number assumes that all massive binaries were formed in a triple, as we assume in our methodology. To eventually account for the possibility of binaries born without a tertiary, one needs only to multiply this number by the fraction of massive binaries in this mass range with a tertiary companion in this mass range. However, observations show that most OB type stars should reside in triple systems \citep{Moe2017,Moe2019}.}. Rate estimates listed in \cite{gwratesreview} for the cluster channel range from $0.2\gpcyr$ to $60\gpcyr$ with most estimates falling roughly within $4\gpcyr$ to $20\gpcyr$. However, there is much variation in the literature in the rates of BBH mergers from both of these channels, especially from the isolated binary evolution channel, arising from uncertainties in the physics of binary stellar evolution \citep[e.g., ][]{2021arXiv210810885B}. Rate estimates have ranged from as low as $0.4\gpcyr$ to $7\times10^3\gpcyr$, with most estimates falling in the range $10\gpcyr$ to $10^3\gpcyr$ \citep[][]{gwratesreview}.

One uncertainty in our work arises from the fitting formula employed by \texttt{COSMIC} to determine stellar radii. \texttt{COSMIC} uses the same fitting formula from \texttt{SSE}, which, as \cite{Agrawal+2020} demonstrates, drastically overestimates the maximum stellar radius of massive stars compared to \texttt{MESA} models (by a factor of 20 in the most extreme cases). However, we expect this to preferentially effect the rate of non-LK mergers since the inner binaries that merge through the LK mechanism tend to be initially wider.

As stated previously, the presence or absence of a common envelope phase plays a large role in shaping the mass ratio distribution of the merging systems. These common envelope phases are important because inner binaries which undergo a common envelope phase typically cannot merge through the LK effect due to GR precession \citep[e.g.,][]{Antonini2012,Rodriguez2018a}. This is confirmed by our results, as while the majority of the merging binary systems experienced a common envelope phase whereas only a single triple system did. However, there are two major uncertainties associated with unstable mass-transfer physics. The first relates to the value of the $\alpha$ parameter used in population synthesis to characterize the efficiency of unbinding the envelope during a common envelope phase. In this study, we used $\alpha=1$. However, \cite{fragos+2019} suggest that values as high as $\alpha=5$ are possible based on newer one-dimensional stellar structure simulations including additional energy sources. Furthermore, recent GW observations suggest that such high efficiencies could be preferred in order to match the astrophysical merger rate implied by observations \citep[][]{giacobbo+mapelli2018,FragioneLoebR2021,santoliquido+2021,zevin+2021}. Previous studies looking at the impact of common envelope phases find that varying the value of this parameter does not strongly change the shape of the mass ratio distribution \citep[][]{giacobbo+mapelli2018,Zevin2020}. We expect that this uncertainty only changes the relative rate of non-LK to LK mergers, bringing the overall non-LK rate down. 

The second uncertainty relates to the onset of stable versus unstable mass transfer. Recently, a study by \cite{olejak+2021} showed that, with more careful prescriptions related to the onset of unstable mass transfer and thermal timescale mass transfer, the rates, mass/mass ratio distribution, and dominant evolution channels of BBH mergers can change drastically. In particular, the main channel that was shown to produce low-$q$ events by \cite{OlejakFishbach2020} was gone with these new prescriptions. Moreover, more detailed work using 1D stellar hydrodynamical codes have shown that current population synthesis schemes may be overpredicting the relative contribution of the common-envelope channel \citep{marchant+2021,gallegosgarcia+2021,klencki+2021}.

There are further uncertainties related to stable mass transfer in these systems. As stated previously, a large fraction of the inner binaries in the KL merger channel underwent at least one phase of stable mass transfer. First of all, it is not clear how accurate this number actually is due to the previously discussed issues with the \texttt{SSE} stellar tracks. Ignoring that, the key uncertainty here is related to the efficiency of these mass transfer episodes. They could be fully conservative (i.e. all the mass from the donor is accreted) or fully non-conservative (i.e. none of the mass from the donor is accreted) or anything in-between. For example, in previous studies using the population synthesis code \texttt{StarTrack}, the code instead defines as a parameter the fraction of mass from the donor that will be accreted, with a default value of $f=0.5$ \citep{Belczynski2008}. Varying the value of this parameter resulted in wider mass ratio distributions for lower values of $f$ \citep[][]{Dominik2012,bel2020}. This result is for two reasons. First, the accretor, which is generally lower mass, simply gains less mass for mass transfer that is less conservative. Second, more mass lost from the system results in further orbital expansion, which can prevent the onset of a common envelope phase. For this study, this also has the further effect of potentially increasing the number of LK mergers relative to non-LK mergers, as non-LK mergers require small separations to efficiently emit GWs. Furthermore, this orbital expansion can increase the number of systems that are not susceptible to quenching of the LK mechanism due to relativistic precession.

Despite the uncertainties in population synthesis models, it has been proven analytically that, at the octupole level of approximation, the LK mechanism preferentially causes the merger of low-$q$ inner binaries \citep{2021arXiv210301963S}. Therefore, once phases of stable and unstable mass transfer have taken the initial population and created a distribution in semi-major axis and mass ratio, the LK mechanism’s preference for low-$q$ mergers means the contribution from LK mergers will typically be higher at low $q$'s than from non-LK mergers. The uncertainties discussed above are unlikely to affect this conclusion, as this effect is completely independent of the details of binary and triple stellar evolution.

In this study, we only looked at the field triples in an isolated environment and so we considered only GWs and the eccentric LK effect as the only dynamical effects leading to mergers. In reality, even field triples could have interactions with passing stars, in some cases leading to merger \citep{michaely+perets2020}. Furthermore, \cite{VignaGomez2021} showed that sufficiently compact triples in the field may still produce some number of sequential BBH mergers. Additionally, \cite{stegmann+2022} find that triples in which the inner binary mergers prior to BH formation, typically during the MS phase, can lead to the formation of low-$q$ mergers similar to this study.

As the number of GW observations continues to grow, it will become much easier to distinguish between different merger formation channels. We have shown the potential for future observations to constrain the importance of hierarchical triples to the cumulative BH merger rate given the efficiency of producing high total mass, low mass ratio events. We leave it to a future study to probe the sensitivity of these conclusions to different assumptions, such as those concerning binary evolution or accounting for mass transfer from the tertiary companion.

\software{\texttt{numpy} \citep{walt_numpy_2011}, \texttt{scipy} \citep{2020SciPy}, \texttt{matplotlib} \citep{hunter_matplotlib:_2007}, \texttt{COSMIC} \citep{2020zndo...3905335B}}

\acknowledgments

We thank Kyle Kremer, Scott Coughlin, Michael Zevin, Max Moe, and Rosanne Di Stefano for useful discussions and suggestions. We thank Kyle Kremer for providing the cluster models used in this work. C.R. acknowledges support from NSF Grant AST-2009916 at Carnegie Mellon University and a New Investigator Research Grant from the Charles E. Kaufman Foundation. G.F.\ acknowledges support from NASA Grant 80NSSC21K1722. Computations were supported in part through the resources and staff contributions provided for the Quest high performance computing facility at Northwestern University, which is jointly supported by the Office of the Provost, the Office for Research, and Northwestern University Information Technology. This work also used computing resources at the Center for Interdisciplinary Exploration and Research in Astrophysics (CIERA) funded by NSF Grant PHY-1726951 and computing resources provided by CIERA.

\bibliographystyle{aasjournal}
\bibliography{refs}

\end{document}